# Localization in Quasi-1D Systems with Random Magnetic Field

Yakov Rutman, Mario Feingold and Yshai Avishai

*Dept. of Physics, Ben-Gurion University, Beer-Sheva 84105, Israel*

**Abstract**

We investigate the localization of electrons hopping on quasi-1D strips in the presence of random magnetic field. In the weak-disorder region, by perturbative analytical techniques, we derive scaling laws for the localization length, $\xi$, of the form $\xi \propto \frac{1}{w^\eta}$, where $w$ is the size of magnetic disorder and the exponent $\eta$ assumes different values in the various energy ranges. Moreover, in the neighborhood of the energies where a new channel opens a certain rearrangement of the perturbation expansion leads to scaling functions for $\xi$. Although the latter are in general quantitatively wrong, they correctly reproduce the corresponding $\eta$ exponents and the form of the scaling variables and are therefore useful for understanding the behavior of $\xi$.

## 1. Introduction.

The problem of electron localization in the presence of a random magnetic field was originally suggested simply as an alternative new type of disorder.[1] Recently however, a realization of the random flux model was proposed by Kalmeyer and Zhang[2] in order to describe the behavior of the Hall system in the neighborhood of fractional filling fractions with even denominator, $\nu = \frac{m}{2n}$. While for such $\nu$'s, the Hall conductance, $\sigma_{xy}$, is not quantized, the corresponding longitudinal conductance, $\sigma_{xx}$, displays a pronounced minimum which is relatively insensitive to changes of the temperature. This suggests the presence of a *Hall metallic phase* that can be explained by replacing the entire magnetic flux with flux tubes attached to the electrons. If for example, $\nu = \frac{1}{2}$, then each electron will have incorporated two flux quanta, leading to a system of composite fermions in zero average magnetic field. Moreover, due to the underlying disorder in the Hall system, the electron density is fluctuating which in turn leads to fluctuations in the associated flux density. Correspondingly, a positive magnetoresistance for random flux models would account for the observed minimum in $\sigma_{xx}$. In turn, the presence of a mobility edge in such models is the simplest mechanism leading to positive magnetoresistance. In a different context, the quantum two-dimensional motion of a particle in a random static magnetic field is also important for the theory of correlated spin system.[3] Sometimes these systems form a spin liquid state and such behavior is likely to occur in high-$T_c$ cuprates.

At first glance, properties of related models do not indicate the possibility of a metal-insulator transition in a 2D random flux model. In particular, in the Anderson model with site disorder alone all states are localized and the corresponding magnetoresistance is negative. Moreover, calculations using supersymmetry methods show that no extended states are to be found in 2D disordered systems with either orthogonal or unitary symmetry and only the symplectic case displays a metal-insulator transition. On the other hand, in the case of 2D disordered systems in a strong constant transverse magnetic field it is known that states at the center of the Landau band are extended. This is due to an additional topological term that appears in the corresponding non-linear sigma model Lagrangian, altering the nature of this system and allowing for the presence of extended states. In the light of recent numerical studies of random flux models that found evidence of extended states,[2,4-7] it appears that this case is also nonrepresentative of its symmetry class.

One way of approximating the 2D random magnetic field problem is via squares of finite size, $L$. It was found[6,7] that for $|E| < E_c \approx 3$, $\xi_L/L$ grows as $L$ is increased while it decreases when $|E| > E_c$. Such behavior indicates that a metal-insulator transition takes place at $E_c$. A different approach to investigating the properties of the infinite size 2D systems uses quasi-1D strips of length $N \to \infty$ and finite width, $M$. Although all states of the quasi-1D system are localized, one expects that knowledge of the behavior of the localization length for such systems can help to elucidate the above mentioned questions. However, numerical results for the random flux model on strips[5,6] are to some extent contradictory. In Ref. 6 the observed behavior is analogous to that obtained for square systems, namely, $\Lambda(M) \equiv \xi_M/M$ changes from an increasing function to a decreasing one at roughly the same value of the energy, $E_c$. On the other hand, in Ref. 5 no evidence of a metal-insulator transition was observed and $\Lambda(M)$ is an increasing function for all the energies that were studied. The apparent disagreement between the two studies could be a result of the fact that, due to numerical limitations, in Ref. 5 only energies larger than 2.95, $|E| > 2.95$, were considered. A more interesting possibility is that the difference between the two studies is related to the short range correlations that are included in the model of Ref. 6 but are absent in that of Ref. 5.

While all numerical studies were done for large magnetic disorder, extended states are certainly present in the limit of vanishing disorder and therefore, one expects that this regime is of interest in the search for a metal-insulator transition in infinitely wide strips. The advantage of the approach where weakly disordered quasi-1D strips are studied is that a transfer matrix formalism can be used for which an appropriate perturbation theory is available.[8] Recently, Avishai and Luck[9] have employed the perturbation theory of Ref. 8 in order to investigate the localization of electrons on a ladder network of two quantum wires with a random magnetic flux on each plaquette. On the other hand, we shall address the corresponding localization problem within the tight-binding approximation and for arbitrary $M$. While in the Anderson model disorder is implemented through random site energies, magnetic disorder is included via random phases in the hopping matrix elements. These phases are usually chosen as independent variables that have the same probability distribution with some definite width, $w$. The purpose of this work is to obtain a perturbative expansion for the localization length, $\xi$, in the limit of weak magnetic disorder.

In Section 2, we present our model and derive the weak-disorder expansion for the positive Lyapounov exponents. The outcome is expressed in several different forms corresponding to the various energy ranges. In Section 3, these predictions are corroborated by accurate numerical data for two wires. Finally, in Section 4, we discuss the results and some unresolved questions left for future work.

## 2. The Model

We study the localization of an electron on quasi-1D strips subject to a random transverse magnetic field. The model we use consists of a tight-binding Hamiltonian with phase disorder in the hopping matrix elements and rigid boundary conditions. The underlying 2D square lattice has unit lattice constant, $a_L = 1$, length $N \to \infty$, $1 \le x \le N$, and width $M$, $1 \le y \le M$. Thus, the lattice sites are located at integer values of $x$ and $y$, $n$ and $m$, respectively. In order to isolate the behavior due to magnetic disorder, in what follows, we assume that the site energies, $\epsilon_{n,m}$, vanish, $\epsilon_{n,m} = 0$. A random magnetic field, $\mathbf{B}$, with zero average is applied parallel to the $z$-axis, $\mathbf{B} = B\hat{z}$. It is chosen such that the corresponding vector potential in the Landau gauge, $\mathbf{A} = (0, Bx, 0)$, on each vertical lattice link, $[n, m]_y$, between the sites $(n, m)$ and $(n, m+1)$, $A_{n,m}$, is an independent random variable. Moreover, all $A_{n,m}$ are identically distributed. Although this is clearly not the most natural choice, it is among the few models for which there is no correlation between consecutive transfer matrices and accordingly, the perturbative expansion of Ref. 8 can be used. Let $|n, m>$ be a complete set of orthogonal states associated with the lattice sites. Then the Hamiltonian takes the form

$$H = \sum_{n,m} |n+1,m><m,n| + |n-1,m><m,n|$$
$$+ e^{i\alpha_{n,m}} |n, m+1><m,n|$$
$$+ e^{-i\alpha_{n,m-1}} |n, m-1><m,n|, \qquad (1)$$

where $\alpha_{n,m} = \frac{e}{\hbar c} A_{n,m}$, and the value of the corresponding eigenfunction at the $(n, m)$ lattice site, $\psi_{n,m}$, satisfies

$$\psi_{n+1,m} + \psi_{n-1,m} + e^{i\alpha_{n,m}} \psi_{n,m+1} + e^{-i\alpha_{n,m-1}} \psi_{n,m-1}$$
$$= E\psi_{n,m}. \qquad (2)$$

Eq. (2) can be written in transfer matrix form. Acting with the $2M \times 2M$ transfer matrix, $T_n$, on the $2M$-vector representing the wave function of two consecutive columns of the strip, $\psi_{n-1,m}$ and $\psi_{n,m}$, where $1 \le m \le M$, generates the vector corresponding to $\psi_{n,m}$ and $\psi_{n+1,m}$. The propagation along the strip is therefore described by the product

$$Q_N = \prod_{n=1}^{N} T_n. \qquad (3)$$

and the corresponding localization length is related to the Lyapounov exponents, $\gamma_i$, of the infinite product $Q = \lim_{N \to \infty} Q_N$. In fact, if $Re\gamma_i \ge Re\gamma_{i+1}$, then $\xi$ is the inverse of the real part of the smallest positive Lyapounov exponent.

In our model, we can write the random transfer matrix $T_n(\alpha)$

$$T_n(\alpha) = \begin{pmatrix} V_n & -I \\ I & O \end{pmatrix} , \quad (4)$$

where $I$, $O$ and $V_n$ are $M \times M$ blocks; $I$ is the corresponding identity matrix, $O$ is the null matrix, and

$$V_n = \begin{pmatrix} E & -e^{i\alpha_{n,1}} & 0 & \cdots \\ -e^{-i\alpha_{n,1}} & E & -e^{i\alpha_{n,2}} & \cdots \\ 0 & -e^{-i\alpha_{n,2}} & E & \cdots \\ \cdots & \cdots & \cdots & \cdots \end{pmatrix} . \quad (5)$$

We assume rigid boundary conditions on the horizontal edges of the strip, $\psi_{n,0} = \psi_{n,M+1} = 0$. Correspondingly, the matrix element situated in the upper right (and lower left) corner of the $V_n$ matrix, vanishes. Moreover, $T_n$ only depends on the vector potential on the $n$-th column of the strip. Since the $\alpha_{n,m}$ are uncorrelated, this implies that the $T_n$ for different $n$ are uncorrelated as well,

$$\overline{T_{n_1} T_{n_2}} = \overline{T}_{n_1} \overline{T}_{n_2} \quad , \text{ for } n_1 \neq n_2 , \quad (6)$$

where $\overline{T}$ is the average of the matrix $T$ over the disorder. This would not be true if, for example, the Landau gauge $\mathbf{A} = (By, 0, 0)$ would have been used. In this gauge, $T_n$ depends on both $A_{n,m}$ and $A_{n-1,m}$ and accordingly, Eq. (6) only holds if $|n_1 - n_2| > 1$.

In the site representation, current conservation implies that the transfer matrix is constrained to satisfy the relation

$$T^\dagger J T = J , \quad (7)$$

where

$$J = \begin{pmatrix} O & -I \\ I & O \end{pmatrix} . \quad (8)$$

For the case where $T_n$ has the structure of Eq. 4, this is quivalent to $V_n$ being hermitian, $V_n = V_n^\dagger$, which is indeed satisfied by the $V_n$ corresponding to our model (see Eq. 5). In general, from Eq. 7 one obtains that the eigenvalues of complex transfer matrices come in pairs, $(i, j)$, $i \neq j$, such that, $\lambda_i = (\lambda_j^*)^{-1}$. Consequently, eigenvalues of complex $T_n$ that lie on the unit circle are at least twofold degenerate.

In the weak-disorder regime, i.e., when the width of the distribution of the random magnetic field is small, the Lyapounov exponents can be expanded in a systematic way in terms of the successive moments of this distribution. This paper presents the perturbative expansion of the Lyapounov exponents and gives explicit expressions for the results up to the second order, following closely the approach described in Ref. 8.

In the case of potential disorder, the corresponding transfer matrix is additively modified by the disorder, that is, $T_n = A + \mu B_n$, where $\mu$ is the width of the distribution of site energies. For magnetic disorder however, this simplifying feature is absent. Nevertheless, one can always obtain an analogous separation using the average transfer matrix, $\overline{T}_n$

$$T_n = \overline{T}_n - (\overline{T}_n - T_n) . \quad (9)$$

Accordingly, the zero order approximation to the Lyapounov exponents is given by the eigenvalues of $\overline{T}_n$, $\lambda_k$, namely, $\gamma_k = \log \lambda_k$. In turn, the $\lambda_k$ satisfy the relation

$$\lambda_k + \lambda_k^{-1} = E - 2a \cos q_k , \quad (10)$$

where

$$q_k = \frac{\pi}{M+1} s_k , \quad (11)$$

and $a = \overline{e^{i\alpha}} = 1 - \frac{\overline{\alpha^2}}{2} + O(\alpha^4)$. Moreover, $s_k$ is a permutation of the integers $(1, 2, ..., M)$, such that

$$|\lambda_1| \geq |\lambda_2| \geq ... \geq |\lambda_k| \geq ... \geq |\lambda_M| . \quad (12)$$

While for $E > 0$, Eq. (12) is satisfied if $s_k = M + 1 - k$, when $E < 0$, it implies that $s_k = k$. Since at $E = 0$, $|\lambda_k| = 1$ for all $k$, the different labeling between the positive and negative energy sectors of the band does not lead to discontinuities in $|\lambda_k|$ at the band center. The real part of $\lambda_k$ however, $\text{Re}\lambda_k$, flips its sign at $E = 0$. One can see from Eq. (10) that the eigenvalues of $\overline{T}$ depend on $a$ and therefore, the energy where the $k$-th channel opens, $E_k$, depends on disorder, $E_k(a)$. It is determined by the requirement that $\lambda_k(a) = \pm 1$, leading to $E_k(a) = 2a \cos q_k \pm 2$. Therefore, the width of the band, $\Delta \equiv 2E_1$, gradually shrinks as the variance of the magnetic disorder is increased. To first order in the variance, $\Delta = 4 + (4 - 2\overline{\alpha^2}) \cos \frac{\pi}{M+1}$.

In order to obtain the higher order terms of the perturbation expansion, one must write the random matrix, $(\overline{T}_n - T_n)$, in the basis where $\overline{T}_n$ is diagonal. For simplicity, we assume that the distribution of $\alpha_{n,m}$ is symmetric and as a consequence, the $O(\overline{\alpha^3})$ term vanishes. Since the $O(\overline{\alpha^4})$ term is rather involved, we only quote here the series up to $O(\overline{\alpha^2})$

$$\sum_{i=1}^p \gamma_i = \sum_{i=1}^p \log \lambda_i - \frac{4\overline{\alpha^2}}{(M+1)^2}$$
$$\times \sum_{i=1}^p \sum_{j=1}^p \frac{T_{ij}}{(\lambda_i(1) - \lambda_i(1)^{-1})(\lambda_j(1) - \lambda_j(1)^{-1})} , \quad (13)$$

where

$$T_{ij} = \sin\left(\frac{q_i}{2}\right) \sum_{k=1}^{M} \sin(kq_i)\sin(kq_j)[\sin((k-1)q_j)$$
$$\times \cos((k-\frac{1}{2})q_i) - \sin((k+1)q_j)\cos((k+\frac{1}{2})q_i)] . \quad (14)$$

It turns out however that the expansion in the moments of $\alpha$ is not well behaved for all energies, $E$. The most serious difficulty originates with the $O(\overline{\alpha^4})$ term that we ignored in Eq. (13). In Ref. 8, it is shown that this term includes sums of ratios of different eigenvalues of $\overline{T}$, e.g.

$$S_{ij} = \frac{1}{N} \sum_{1 \leq \alpha < \beta \leq N} \left(\frac{\lambda_i}{\lambda_j}\right)^{\beta-\alpha} , \quad (15)$$

where $p < i \leq M$ and $1 \leq j \leq p$. In the limit $N \to \infty$, $S_{ij}$ does not converge whenever $|\lambda_i| = |\lambda_j| = 1$ and for energies within the band there always exists some $(i,j)$ pair for which that is the case. The only exception occurs for the energy range in which there is a single open channel, $2a\cos\frac{2\pi}{M+1} + 2 < |E| < 2a\cos\frac{\pi}{M+1} + 2$, where the $O(\overline{\alpha^4})$ term is convergent despite the fact that $|\lambda_M| = |\lambda_{M+1}| = 1$. Another type of divergence occurs in the $O(\overline{\alpha^2})$ term at the energies at which $\lambda_k(1) = \pm 1$ where $1 \leq k \leq p$ (see Eq. (13)), $\tilde{E}_k = 2\cos q_k \pm 2$. While in the case of potential disorder, these divergences occur whenever a new channel opens, $\tilde{E}_k = E_k$, here these do not coincide, $E_k - \tilde{E}_k = 2(a-1)\cos q_k = O(\overline{\alpha^2})$. In order to avoid both these divergences, it is enough that a strong version of Eq. (12), where all the weak inequalities, $\geq$, are replaced by strong ones, $>$, be satisfied. In the following we shall refer to this requirement as the *nondegeneracy condition*. On the other hand, such condition excludes most of the energy band which is the range of physical interest. Two ways have been suggested to obtain information on the Lyapounov exponents using the $\alpha$-expansion in the range where the nondegeneracy condition fails. In Ref. 8, the energy is allowed to be complex, $E_D = E + i\epsilon$, which for a fixed, nonvanishing $\epsilon$ insures that the nondegeneracy condition is satisfied everywhere except for a finite number of isolated points. However, since it is not clear how to extrapolate the results down to the real energy axis, this approach, while conceptually promising, is merely an uncontrolled approximation. On the other hand, in previous work,[11] we have shown that in the neighborhood of the singularities of the $O(\overline{\alpha^2})$ term, $\tilde{E}_k$, useful information on the localization length can be obtained by ignoring the ill-behaved $O(\overline{\alpha^4})$ term. In other words, certain properties of $\xi$ can be extracted from the $\alpha$-expansion up to second order and are not modified by the higher order terms. Specifically, for small $t \equiv E - \tilde{E}_k$ and $w^2 \equiv \overline{\alpha^2}$, the expansion can be written in terms of new variables, $x_i = w^{-\beta_i} t^{\tau_i}$, that balance the largeness of the divergent terms with the smallness of disorder. We refer to $x_i$ as scaling variables and to the divergent terms as resonant. From the structure of the $\alpha$-expansion one can obtain the $\beta_i$ and $\tau_i$ exponents.[10,11] Moreover, keeping only the resonant terms, the perturbation expansion is rearranged such that

$$\gamma_M = w^{\eta_1} f_{\eta_1}(x_1) + w^{\eta_2} f_{\eta_2}(x_2) + \ldots , \quad (16)$$

where $\eta_1 < \eta_2 < \ldots < \eta_i < \ldots$, and $f_{\eta_i}$ are non-divergent scaling functions. Since $w$ is small, we naturally restrict our analysis to the first term of Eq. (16). Moreover, we find the asymptotic form of $f_{\eta_1}(x_1)$ from the requirement that for large $x_1$ it matches up with the resonant part of the original perturbation expansion.

Quantitatively however the prediction of Eq. (16) is false, although in some cases the corresponding error is extremely small, e.g. for potential disorder with $M = 3$ and $k = 2$. We also noticed that the value of the theoretical $O(\overline{\alpha^2})$ term turns out to be larger than the correct value in the several cases that were studied. One is therefore tempted to conjecture that the truncated $\alpha$-expansion leads to a lower bound, $\xi_T$, for the true localization length. However, for potential disorder this implies that states are extended in the corresponding two-dimensional case[12] which is known to be wrong.

We now consider the scaling approach to Eq. (13) for the $M > 1$ case of the random field model. In the neighborhood of the outermost singularity, $E = \tilde{E}_1$, we find that $\eta_1 = 1$, $\beta_1 = 2$, $\tau_1 = 1$ and from Eqs. (13) - (14)

$$f_{\eta_1}(x) \simeq \sqrt{x} + \frac{1}{12\sqrt{x}}\Big(1 - \frac{4}{(M+1)^2}$$
$$\times \sum_{j=1}^{M-1} \frac{(T_{Mj} + T_{jM})}{\sqrt{(1 + 2\cos\frac{\pi(1+j)}{2(M+1)}\cos\frac{\pi(1-j)}{2(M+1)})^2 - 1}}\Big) . \quad (17)$$

On the other hand, for $\tilde{E}_i$ with $i > 1$, $\eta_1 = 4/3$, $\beta_1 = 4/3$, $\tau_1 = 1$. The real part of the corresponding scaling function is for $x < 0$

$$\mathrm{Re} f_{\eta_{4/3}}(x)$$
$$\simeq \frac{(T_{Mi} + T_{iM})}{3(M+1)^2\sqrt{-x}\sqrt{\left(1 - 2\cos\frac{\pi(1+i)}{2(M+1)}\cos\frac{\pi(1-i)}{2(M+1)}\right)^2 - 1}} . \quad (18)$$

For $x > 0$ however, all the resonant terms in Eq. (13) are imaginary.

### 3. Comparison with numerical simulations.

In order to compare the theoretical predictions with the results of numerical simulation, we consider a strip of width $M = 2$ and a uniform distribution for the vector potentials, $\alpha_{n,m}$,

$$P(\alpha_{n,m}) = \begin{cases} \frac{1}{2w_1}, & \text{for } |\alpha_{n,m}| < w_1 \\ 0, & \text{for } |\alpha_{n,m}| > w_1 \end{cases} \quad (19)$$

where $w_1 = \sqrt{3}w$. The spectrum of the ordered system consists of three domains, namely, I. $|E| > 3$, II. $1 < |E| \leq 3$ and III. $|E| \leq 1$. Their edges are defined by the $\lambda_2(1) = \pm 1$ and $\lambda_1(1) = \pm 1$ conditions, respectively and the localization length has different forms in each of these regions. Since the behavior of the localization length is invariant under $E \to -E$, in what follows, we only consider the positive energy range, $E > 0$.

In the region $1 < E \leq 3$, one of the $\lambda$-eigenvalues is real, $\lambda_1$, and the other, $\lambda_2$, is imaginary. In other words, only one transversal channel is open. In order to obtain $\gamma_2$, and hence the localization length, we have used the perturbative result of Eq. (13). Since $\lambda_2$ is imaginary, neither the term of order zero nor the term proportional to $w^2$ contribute to $\gamma_2$. As a consequence, only fourth-order terms contribute to $\gamma_2$ and therefore, the inverse localization length vanishes as $w^4$. Indeed, in Fig. 1 the inverse localization length at $E = 2.5$ is shown to grow as $w_1^\kappa$ with $\kappa = 4.00 \pm 0.03$, in agreement with our prediction.

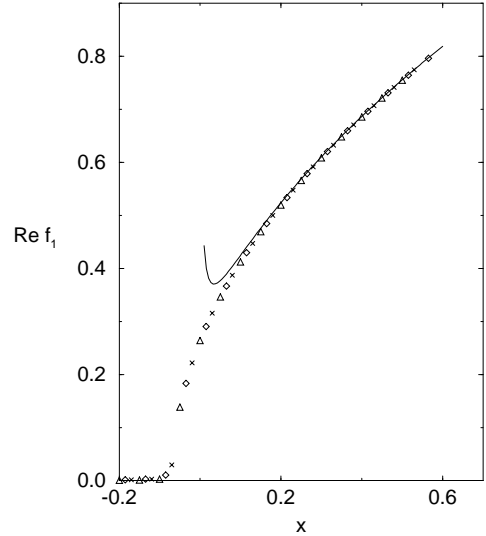

Fig. 2. The real part of the scaling function $f_1(x)$ describing the behavior of smallest Lyapounov exponent $\gamma_2$ near external band edge, $E = 3$. The data correspond to rectangular distributions of the random values, with various values of the width, $w_1 = 0.1$ ($\triangle$), 0.15 ($\times$) and 0.2 ($\diamond$). The continuous line represents the scaling curve given by Eq. (17).

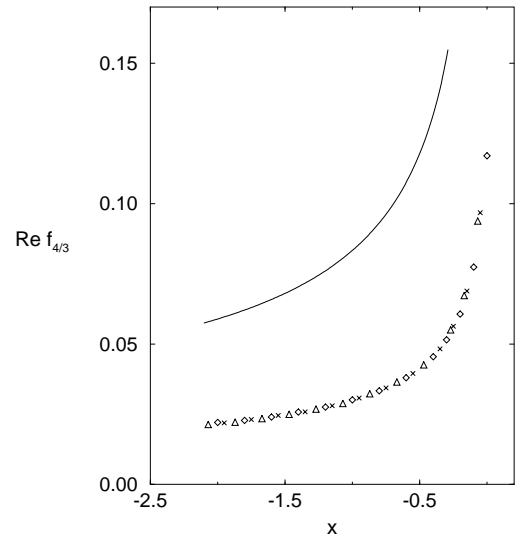

Fig. 3. Same as in Fig. 2 only near internal band edge, $E = 1$. The continuous line represents the scaling curve given by Eq. (18).

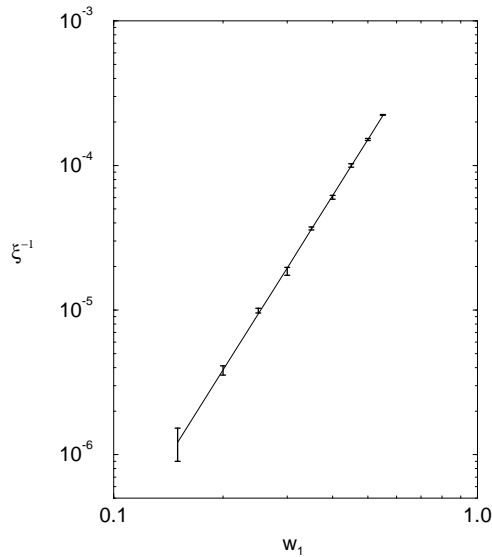

Fig. 1. Numerical values of the inverse localization length, $\xi^{-1}$, as a function of the strength of disorder, $w_1$, at $E = 2.5$ (error bars). The straight line is the best fit of the form $\xi^{-1} = Cw_1^\kappa$, where $\kappa = 4.00 \pm 0.03$.

For energy $E < 1$, both eigenvalues, $\lambda_1$ and $\lambda_2$, are imaginary, i.e. both transversal channels are open. Un-fortunately, the general weak-disorder perturbative result of Eq. (13) does not yield directly any quantitative information about the Lyapounov exponents in this region. While for large enough $\epsilon$ the perturbation theory is valid, such addition to the energy changes the physics of the problem; the perturbation theory prediction now refers to the $\xi$ corresponding to a physically irrelevant model with complex energy that significantly differs from the one with real energy. On the other hand, if the minimal $\epsilon$ required for regularizing the perturbation expansion, $\epsilon_c$, is not too large, one could hope that $\xi(\epsilon_c)$ is a good estimate of $\xi(0)$. Unfortunately, numerical simulation shows that this is not the case. Consequently, both Lyapounov exponents vanish like $w^2$, but one cannot obtain explicit expressions for the amplitude functions.

We now look at the behavior of the second Lyapounov exponent, $\gamma_2$, near the external band edge of the spectrum of the unperturbed network, that is, for $E \to 3$. For $x \to +\infty$, that is, deep outside the band, the result for the scaling law can be written in the form

$$\gamma_2 = w(\sqrt{x} + \frac{1 - \frac{1}{\sqrt{3}}}{12\sqrt{x}}) \qquad (20)$$

where $x = \frac{E-3}{w^2}$. Conversely, for x $\to -\infty$, i.e., deep in the region with a single open channel the second Lyapounov exponent vanishes as the fourth power of the strength of disorder. Fig. 2 shows the real part of the scaling function $f_1 = \frac{\gamma_2}{w}$, obtained from data corresponding to a narrow distribution of vector potentials. For $x > 0$, scaling is observed to hold extremely well and so does Eq. (20) for large enough values of $x$. On the other hand, for negative $x$, scaling holds only in a small interval, up to $x \approx -0.05$.

For $E \to 1$, the spectrum of the network exhibits an internal band edge, which demarcates the two-channel region from the one-channel region. The perturbative weak-disorder expansion of Eq. (13) for the second Lyapounov exponent is again singular. In analogy with the previous case, we are led to hypothesize the scaling form

$$\gamma_2 = w^{4/3} \frac{1}{12\sqrt{-x}} \qquad \text{for} \qquad x \to -\infty , \qquad (21)$$

where $x = \frac{E-1}{w^{4/3}}$ is the corresponding scaling variable. For $x > 0$ however, all the resonant terms in Eq. (13) are imaginary and one cannot obtain information about this part of the scaling function from the $O(\overline{\alpha^2})$-terms. Fig. 3 shows the real part of the appropriate scaling function, $f_{4/3} = \frac{\gamma_2}{w^{4/3}}$. While scaling holds here as well as it does in Fig. 2, unlike in Fig. 2, the theoretical scaling function is significantly larger than the numerically obtained one. This is a consequence of the failure of the $\alpha$-expansion in this range.

## 4. Conclusions

We have investigated the localization of electrons in the framework of a tight-binding Hamiltonian on quasi-1D strips with a random magnetic field. The magnetic vector potentials are assumed to be independent and drawn from a common even distribution. As in any model of 1D disordered wires, all eigenstates are exponentially localized. We have therefore focussed our attention on the weak-disorder regime, $w \ll 1$, where the localization length, $\xi = 1/\mathrm{Re}\gamma_M$, is much larger than the lattice spacing, $\xi \gg a_L$. In this regime, a systematic perturbative expansion for the smallest positive Lyapounov exponent was derived. Moreover, we have checked the outcome of our analytical approach against data obtained by means of numerical simulations for $M = 2$.

The most important result is that magnetic disorder modifies the divergence laws of the localization length $\xi$ in the weak-disorder regime ($w \to 0$), in the various energy domains, with respect to those observed in the case of potential disorder. If we go continuously from outside the band toward its interior, we encounter the following sequence of exponents for the divergence law of the localization length, $\xi \propto 1/w^\eta$,

$$\eta = 0, 1, 4, 4/3, 2, 4/3, 2, .... \qquad (22)$$

This sequence is more diverse than the one observed for potential disorder[11], namely

$$\eta = 0, 2/3, 2, 4/3, 2, 4/3, 2, .... \qquad (23)$$

We suggest that it should be interesting to use the approach described in this work to study the combined effects of potential and magnetic disorder.[13] Moreover, it would be useful to generalize the theory of Ref. 8 for the case where transfer matrices corresponding to neighboring strip columns are not uncorrelated and find the influence of such correlations on the $\eta$-exponents.

However, the most important task remains finding the way to properly regularize the perturbation expansion for infinite products of transfer matrices such that it would become quantitatively correct throughout the band. This problem is reminiscent of analogous difficulties arising in the Hamiltonian and Green function perturbation theories and one expects that some of the methods developed there could be applied to the case of infinite products as well.

## 5. Acknowledgments

We would like to thank D.P. Arovas, S. Fishman, J.L. Pichard and O. Piro for useful discussions. This work was supported by the Israel Science Foundation administered by the Israel Academy of Sciences and Humanities.